\documentstyle[prb,aps,twocolumn,epsfig]{revtex}

\newcommand{\be}{\begin{eqnarray}}
\newcommand{\ee}{\end{eqnarray}}

\begin{document}
\draft
\twocolumn[\hsize\textwidth\columnwidth\hsize\csname @twocolumnfalse\endcsname
\title{Melting of Charge/Orbital Ordered States in Nd$_{1/2}$Sr$_{1/2}$MnO$_3$:
Temperature and Magnetic Field Dependent Optical Studies}
\author{J. H. Jung, H. J. Lee, and T. W. Noh}
\address{Department of Physics and Center for Strongly Correlated Materials Research, 
\\
Seoul National University, Seoul 151-742, Korea}
\author{E. J. Choi}
\address{Department of Physics, University of Seoul, Seoul 130-743, Korea}
\author{Y. Moritomo}
\address{CIRSE, Nagoya University, Nagoya 464-8603 and PRESTO, JST, Japan}
\author{Y. J. Wang and X. Wei}
\address{National High Magnetic Field Laboratory at Florida State University\\
Tallahassee, FL 32310, USA}
\date{\today }
\maketitle

\begin{abstract}
We investigated the temperature ($T=$ 15 $\sim $ 290 K) and the magnetic
field ($H=$ 0 $\sim $ 17 T) dependent optical conductivity spectra of a
charge/orbital ordered manganite, Nd$_{1/2}$Sr$_{1/2}$MnO$_3$. With
variation of $T$ and $H$, large spectral weight changes were observed up to
4.0 eV. These spectral weight changes could be explained using the polaron
picture. Interestingly, our results suggested that some local ordered state
might remain above the charge ordering temperature, and that the
charge/orbital melted state at a high magnetic field (i.e. at $H=$ 17 T and $%
T=$ 4.2 K) should be a three dimensional ferromagnetic metal. We also
investigated the first order phase transition from the charge/orbital
ordered state to ferromagnetic metallic state using the $T$- and $H$%
-dependent dielectric constants $\varepsilon _1$. In the charge/orbital
ordered insulating state, $\varepsilon _1$ was positive and $d\varepsilon
_1/d\omega \approx 0$. With increasing $T$ and $H$, $\varepsilon _1$ was
increased up to the insulator-metal phase boundaries. And then, $\varepsilon
_1$ abruptly changed into negative and $d\varepsilon _1/d\omega >0$, which
was consistent with typical responses of a metal. Through the analysis of $%
\varepsilon _1$ using an effective medium approximation, we found that the
melting of charge/orbital ordered states should occur through the
percolation of ferromagnetic metal domains.
\end{abstract}


% \twocolumn[\hsize\textwidth\columnwidth\hsize\csname @twocolumnfalse\endcsname

\vskip1pc] \newpage

\section{INTRODUCTION}

Doped manganites, with chemical formula {\it R}$_{1-x}${\it A}$_x$MnO$_3$ [%
{\it R}= La, Nd, Pr, and {\it A}= Ca, Sr, Ba], have attracted lots of
attention due to their exotic transport and magnetic properties, such as
colossal magnetoresistance.\cite{jin} The coexistence of ferromagnetism and
metallicity, for the samples near $x$ $\sim $ 0.3, had been explained by the
double exchange model.\cite{zener} However, it was found that the double
exchange interaction alone cannot explain the colossal magnetoresistance.%
\cite{millis} Additional mechanisms were proposed. Among them, two scenarios
attracted most of attention: the polaron due to the Jahn-Teller distortion
of Mn$^{3+}$ ion\cite{millis,roder} and the orbital fluctuation.\cite
{maekawa,horsch}

On the other hand, some manganite samples with small bandwidths near $x$ $%
\sim $ 1/2 show intriguing charge ordering phenomena,\cite{cheong} i.e. real
space orderings of the Mn$^{3+}$ and the Mn$^{4+}$ ions. For manganites, the
charge ordering is usually accompanied with orbital and antiferromagnetic
ordering. For example, charge ordering in Nd$_{1/2}$Sr$_{1/2}$MnO$_3$ leads
to the {\it d}$_{3{\it x}^2-{\it r}^2}$ ({\it d}$_{3{\it y}^2-{\it r}^2}$)
orbital ordering and the {\it CE}-type antiferromagnetic spin ordering at a
low temperature.\cite{kuwahara} Moreover, it was found that some charge
ordered states could be changed into ferromagnetic metallic states at a
higher temperature and/or under a high magnetic field.\cite{tomioka} The
transitions from charge/orbital ordered insulator to ferromagnetic metal are
usually called ''melting'' of charge/orbital ordered states.

There have been numerous optical investigations which tried to understand
basic mechanisms of colossal magnetoresistance.\cite{oki,quijada,kim98}
However, only a few works have been reported for optical responses of the
charge/orbital ordered state.\cite{liu,calvani,jung_214} Recently, Okimoto 
{\it et al}. reported the magnetic field dependent optical conductivity for
a charge/orbital ordered manganite, Pr$_{0.6}$Ca$_{0.4}$MnO$_3$, and that
the optical responses under the magnetic field could be understood
qualitatively in terms of an insulator-metal transition.\cite{okimoto}
However, their measured spectral region was rather limited (i.e., from
mid-infrared to visible), so details of the insulator-metal transition could
not be addressed.

In this paper, we will report optical properties of Nd$_{1/2}$Sr$_{1/2}$MnO$%
_3$. To get clear understanding on the insulator-metal transitions due to
the melting of the charge/orbital ordered states, optical spectra were taken
by varying either temperature ($T$) or magnetic field ($H$). Our
experimental data will be analyzed in terms of the polaron scenario. The
changes of the optical response due to the melting of the charge/orbital
ordered states will be explained in terms of the percolation model.

\section{EXPERIMENTAL}

Nd$_{1/2}$Sr$_{1/2}$MnO$_3$ single crystal was grown by the floating zone
methods. Details of sample growth and characterization were reported
elsewhere.\cite{moritomo97} The $T$-dependent resistivity was measured by
the four-probe method and the magnetoresistance was obtained using the 20 T
superconducting magnet. For optical measurements, the crystal was polished
up to 0.3 $\mu $m using the diamond paste. To remove surface damages due to
the polishing process, we carefully annealed the sample again in an O$_2$
atmosphere at 1000 $^oC$ just before optical measurements.\cite{haeja}

Near normal incident reflectivity spectra were measured from 5 meV to 30 eV.%
\cite{jung97} A Fourier transform spectrophotometer was used for 5 meV $\sim 
$ 0.8 eV, and a grating monochromator was used for 0.6 $\sim $ 7.0 eV. Above
6 eV, we used the synchrotron radiation from the Normal Incidence
Monochromator beam line at the Pohang Light Source. After the spectra were
taken, the gold normalization technique was used to subtract surface
scattering effects. In the frequency region of 5 meV $\sim $ 4 eV, the $T$%
-dependent reflectivity spectra were taken using the liquid-He cooled
cryostat. In the same frequency region, the $H$-dependent reflectivity
spectra were taken with spectrophotometers at National High Magnetic Field
Laboratory.

The Kramers-Kronig analyses were used to obtain $T$- and $H$-dependent
optical conductivity spectra $\sigma (\omega )$. For these analyses, the
room temperature reflectivity spectrum in the frequency region of 4 $\sim $
30 eV was smoothly connected. Then, the reflectivity at 30 eV was extended
up to 40 eV, above which $\omega ^{-4}$ dependence was assumed. In the low
frequency region, the reflectivity spectrum below 5 meV was extrapolated to
be a constant for an insulating state or using the Hagen-Rubens relation for
a metallic state.\cite{kim97} To check the phase errors due to the
extrapolations in the Kramers-Kronig analyses, we also independently
measured optical constants in the frequency region of 1.5 $\sim $ 5 eV using
a spectroscopic ellipsometry. It was found that the data from the
spectroscopic ellipsometry measurements agreed quite well with the
Kramers-Kronig analyses results, demonstrating the validity of our
extrapolations.

\section{DATA AND\ RESULTS}

\subsection{{\it dc} resistivity}

\begin{figure}[tbp]
\epsfig{file=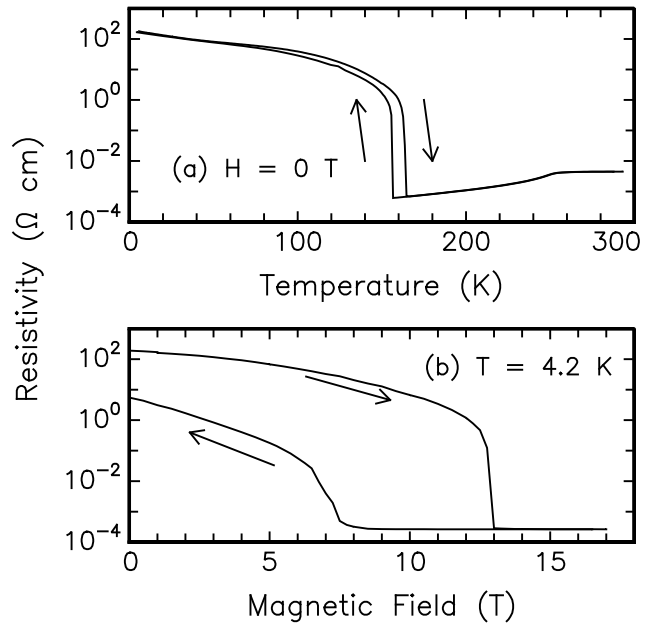,width=3.3in,clip=}
\vspace{2mm}
\caption{(a) $T$- and (b) $H$-dependent {\it dc} resistivity of Nd$_{1/2}$Sr$%
_{1/2}$MnO$_3$. }
\label{Fig:1}
\end{figure}

Figure 1(a) shows the $T$-dependent {\it dc} resistivity curve of Nd$_{1/2}$%
Sr$_{1/2}$MnO$_3$ which was taken with $H$ = 0 T. With decreasing $T$, the 
{\it dc} resistivity value slightly decreases near the ferromagnetic
ordering temperature $T_C$ ($\sim $ 250 K), but it increases abruptly near
the charge ordering temperature $T_{CO}$ ($\sim $ 150 K). The {\it dc}
resistivity value at 4.2 K is estimated to be around 200 $\Omega $ cm. This
large value of the {\it dc} resistivity is known to be originated from real
space charge/orbital ordering. With increasing $T$, the {\it dc} resistivity
value smoothly decreases initially and then experiences an abrupt decrease
to $\sim $ 0.6 $m\Omega $ cm near 170 K. The {\it dc} resistivity values for
the heating run are larger than those for the cooling run, suggesting that
the melting of the charge/orbital ordered states has the nature of first
order phase transition. Above 170 K, the {\it dc} resistivity values are
nearly the same as those for the cooling run. Note that no apparent
hysteresis can be observed near $T_C$.

Figure 1(b) shows the $H$-dependent {\it dc} resistivity curve of Nd$_{1/2}$%
Sr$_{1/2}$MnO$_3$ which was taken at 4.2 K. With increasing $H$, the {\it dc}
resistivity value slowly decreases initially, but it suddenly decreases to $%
\sim $ 0.2 $m\Omega $ cm near 13 T. Above 13 T, the {\it dc} resistivity
value doesn't change at all within our experimental errors. With decreasing $%
H$, the {\it dc} value does not change down to 7.5 T and starts to increase
abruptly near 7.5 T. The {\it dc} resistivity curve shows a very strong
hysteresis below 13 T: the {\it dc} resistivity values for the
field-decreasing run are quite smaller than those for the field-increasing
run. Note that the {\it dc} resistivity value ($\sim $ 0.2 $m\Omega $ cm)
for the ferromagnetic metal state at $H=$ 17 T is lower than that ($\sim $
0.6 $m\Omega $ cm) for the same state at 170 K.

\subsection{Temperature dependent optical conductivity spectra}

\begin{figure}[tbp]
\epsfig{file=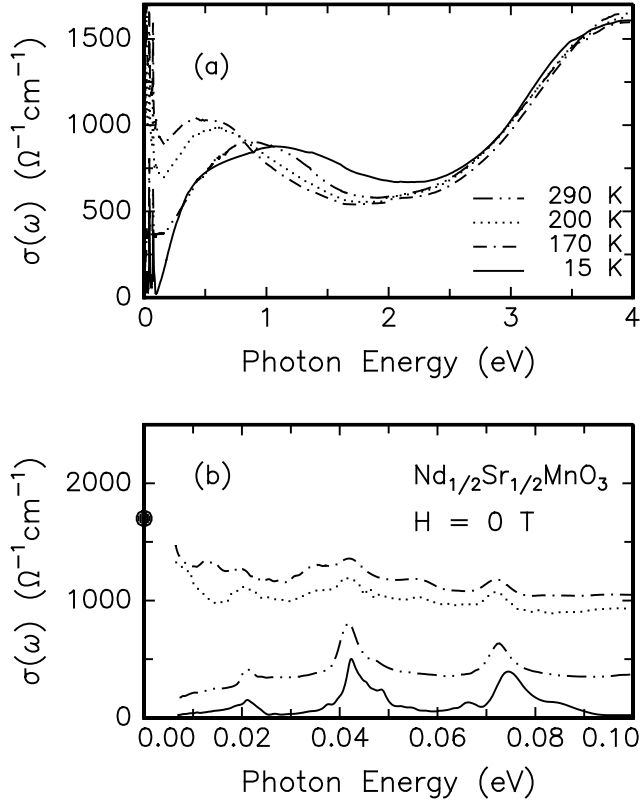,width=3.3in,clip=}
\vspace{2mm}
\caption{$T$-dependent $\sigma (\omega )$ of Nd$_{1/2}$Sr$_{1/2}$MnO$_3$
below (a) 4.0 eV and (b) 0.1 eV. In (b), the solid circle represents the 
{\it dc} conductivity value at 170 K. }
\label{Fig:2}
\end{figure}

The $T$-dependent $\sigma (\omega )$ of Nd$_{1/2}$Sr$_{1/2}$MnO$_3$ are
shown in Fig. 2(a). At room temperature (i.e. $T>T_C$), there are two broad
peaks near 1.0 and 4.0 eV. When entering into the ferromagnetic metallic
state (i.e. $T<T_C$), the broad 1.0 eV peak shifts to a lower energy, which
accompanies large spectral weight changes. In addition, there is a small
decrease of the spectral weight near 3.0 eV. Interestingly, even at a highly
metallic state near 170 K, optical conductivity decreases below 0.5 eV and
shows the Drude-like behavior below 0.1 eV. When entering into the
charge/orbital ordered state (i.e. $T<T_{CO}$), the spectral weights move to
the opposite direction: namely, from a low to a high energy region. The
optical conductivity spectrum at this charge/orbital ordered state shows an
opening of optical gap, whose value is estimated to be about 0.1 eV. [This
value is in reasonable agreements with the value obtained from recent
photoemission experiments.\cite{sekiyama}] In addition, the spectral weights
near 3.0 eV are restored approximately to the values at $T>T_C$.

The far-infrared $\sigma (\omega )$ are displayed in Fig. 2(b). Above $T_C$,
there are three optical phonon peaks, which are known as the external, the
bending, and the stretching modes of the cubic perovskite.\cite{kim96} In
the temperature region of $T_{CO}<T<T_C$, the phonon features are screened
and $\sigma (\omega )$ increase significantly. [The solid circle represents
the {\it dc} conductivity value at 170 K.] Note that the Drude-like
absorption behavior is not so clear.\cite{saitoh} Below $T_{CO}$, $\sigma
(\omega )$ decrease quite drastically. At this low temperature, the bending
and the stretching modes are splitted and corresponding phonon frequencies
move to higher energies. Such changes in the phonon spectra can be
understood in terms of the strong lattice distortion due to the
charge/orbital ordering.\cite{ishikawa}

\subsection{Magnetic field dependent optical conductivity spectra}

\begin{figure}[tbp]
\epsfig{file=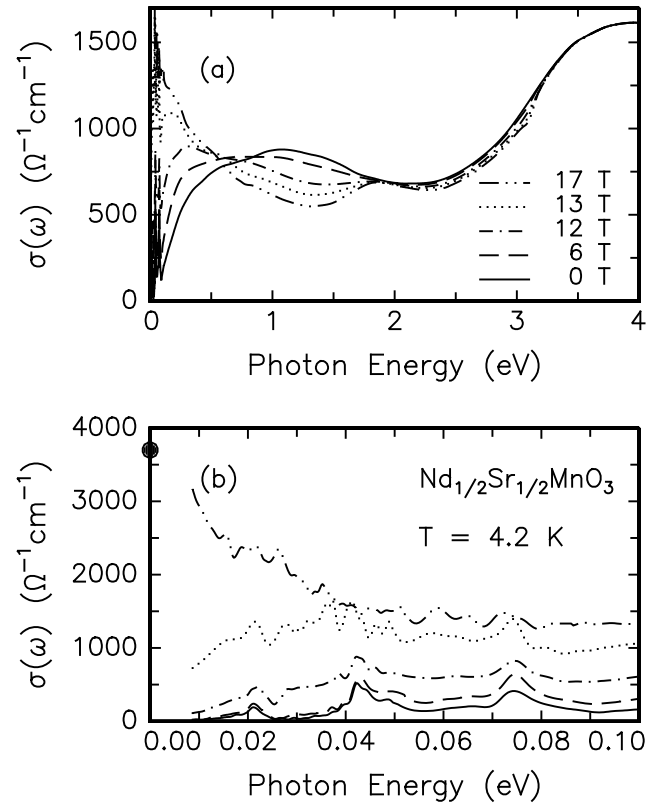,width=3.3in,clip=}
\vspace{2mm}
\caption{$H$-dependent $\sigma (\omega )$ of Nd$_{1/2}$Sr$_{1/2}$MnO$_3$
below (a) 4.0 eV and (b) 0.1 eV. In (b), the solid circle represents the 
{\it dc} conductivity value at 17 T. }
\label{Fig:3}
\end{figure}

The $H$-dependent $\sigma (\omega )$, which were taken at 4.2 K, are shown
in Fig. 3(a). [Note that the spectra were measured with increasing $H$.] At
0 T, the optical spectra are nearly the same as those at 15 K, displayed in
Fig. 2(a). With increasing $H$, the spectral weights near 1.2 and 2.7 eV are
transferred to lower energy regions. The gap values seem to decrease and
finally disappear above 13 T. Note that the $H$-dependent spectral weight
changes are similar to the $T$-dependent spectral weight changes near $%
T_{CO} $. However, the spectra in the ferromagnetic metallic state of 17 T
clearly show a Drude-like absorption feature, which is somewhat different
from $\sigma (\omega )$ of the ferromagnetic metallic state at 170 K,
displayed in Fig. 2(a).

The far-infrared $\sigma (\omega )$ under various $H$ are displayed in Fig.
3(b). At 0 T, the low temperature phonons can be seen clearly. With
increasing $H$, the phonon peaks become screened and the Drude peak seems to
appear above 13 T. [The solid circle represents the {\it dc} conductivity
value at 17 T.] Note that the Drude peak becomes clear and appears below
0.04 eV.

\section{DISCUSSIONS}

\subsection{A schematic diagram of optical transitions}

\begin{figure}[tbp]
\epsfig{file=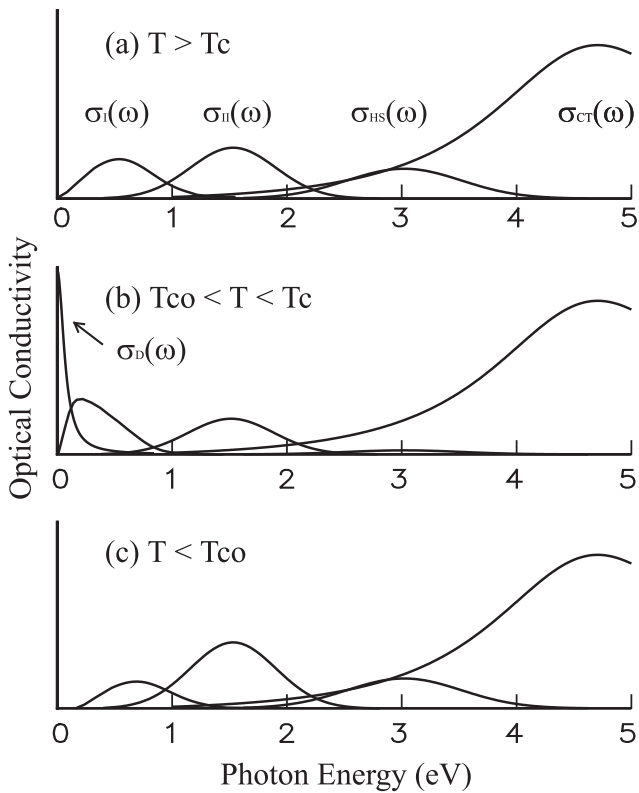,width=3.3in,clip=}
\vspace{2mm}
\caption{Schematic diagram of optical transitions for (a) $T>T_C$, (b) $%
T_{CO}<T<T_C$, and (c) $T<T_{CO}$.}
\label{Fig:4}
\end{figure}

Interpretations on $\sigma (\omega )$ of perovskite manganites have been
quite different among experimental groups.\cite{noh} However, a correct
interpretation is essential to understand physics of the colossal
magnetoresistance and the charge/orbital ordering phenomena. Recently, we
proposed a schematic diagram of $\sigma (\omega )$ based on the polaron
scenario,\cite{haeja,noh} which seems to explain most features of optical
transitions in colossal magnetoresistance manganites observed by numerous
group.\cite{orbital} We want to extend the schematic diagram to include the
charge/orbital ordered state.

Figure 4 shows our proposed schematic diagram of optical transitions in Nd$%
_{1/2}$Sr$_{1/2}$MnO$_3$: (a) $T>T_C$, (b) $T_{CO}<T<T_C$, and (c) $T<T_{CO}$%
. Above $T_C$, there are four main contributions for $\sigma (\omega )$
below 5 eV: (i) $\sigma _I(\omega )$ due to a small polaron absorption below
1.0 eV, (ii) $\sigma _{II}(\omega )$ due to an inter-orbital transition
between the Jahn-Teller splitted levels of the Mn$^{3+}$ ions near 1.5 eV,
(iii) $\sigma _{HS}(\omega )$ due to an optical transition between the
Hund's rule split bands near 3.0 eV, and (iv) $\sigma _{CT}(\omega )$ due to
a charge transfer transition from the O 2{\it p} band to the Mn 3{\it d}
band near 4.5 eV. The optical transition between the Hund's rule split bands
represents {\it e}$_g^{\uparrow }$({\it t}$_{2g}^{\uparrow }$) $\rightarrow $
{\it e}$_g^{\uparrow }$({\it t}$_{2g}^{\downarrow }$) and {\it e}$%
_g^{\downarrow }$({\it t}$_{2g}^{\downarrow }$) $\rightarrow $ {\it e}$%
_g^{\downarrow }$({\it t}$_{2g}^{\uparrow }$) transitions. [This notation
indicates the transition occurs between two {\it e}$_g$ bands with the same
spin but different {\it t}$_{2g}$ spin background.] There have been numerous
optical reports which support our assignments of the small polaron peak,\cite
{kaplan,jung98} the optical transition between Hund's rule split bands,\cite
{moritomo98,jung99} and charge transfer peaks.\cite
{oki,quijada,jung97,jung98} On the other hand, the existence of peak near
1.5 eV was observed by many workers,\cite{quijada,jung98,machida} but there
remain some controversies for its origin.\cite{discrepancy} We think that
the most probable candidate is the inter-orbital transition at the same Mn$%
^{3+}$ site. Although this transition is prohibited by the selection rule
for an Mn atom, this transition could become possible due to the local
lattice distortion of MnO$_6$ octahedra and the strong hybridization between
Mn 3{\it d} and O 2{\it p} orbitals.

For the ferromagnetic metallic region of $T_{CO}<T<T_C$, the small polaron
peak will change into coherent and incoherent absorptions of a large polaron.%
\cite{kim98,yoon} The coherent absorption will appear as Drude-like optical
conductivity spectra $\sigma _D(\omega )$ and the incoherent one as an
asymmetric mid-infrared peak. The increase of electron screening and the
decrease of lattice distortion in the metallic state will decrease the 1.5
eV peak somewhat. The optical transition between the Hund's rule splitted
bands will decrease, since all of the {\it t}$_{2g}$ spins will be aligned
in the ferromagnetic state. And, the charge transfer peak remains to be
nearly $T$-independent.

Below $T_{CO}$, the coherent absorption of the free carrier will disappear
due to the charge/orbital ordering. And, its spectrum will be similar to
that for $T>T_C$. However, there seems to be three minor but important
differences. First, the optical gap due to charge/orbital ordering should
appear. Second, the absorption peak due to the polaron hopping should
decrease since such a hopping requires more energy in the antiferromagnetic
ordered state. Third, the 1.5 eV peak should become stronger, since lattice
distortion becomes larger in the charge/orbital ordered state.

\subsection{Temperature dependent spectral weight changes}

\begin{figure}[tbp]
\epsfig{file=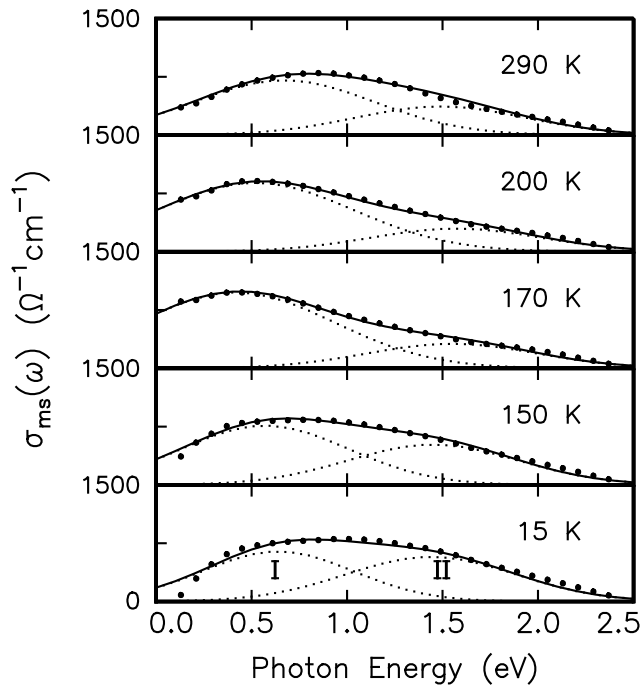,width=3.3in,clip=}
\vspace{2mm}
\caption{$T$-dependent midgap states of Nd$_{1/2}$Sr$_{1/2}$MnO$_3$. The
solid circles, the dotted lines, and the solid lines represent the
experimental data, the Gaussian functions, and the sums of two Gaussian
functions, respectively.}
\label{Fig:5}
\end{figure}

For the quantitative analysis of $T$-dependent electronic structure, we
analyzed $\sigma (\omega )$ in terms of five peaks, discussed in Section IV.
A:

\begin{equation}
\sigma (\omega )=\sigma _D(\omega )+\sigma _I(\omega )+\sigma _{II}(\omega
)+\sigma _{HS}(\omega )+\sigma _{CT}(\omega )\text{ .}
\end{equation}
For $\sigma _{HS}(\omega )$ and $\sigma _{CT}(\omega )$, the Lorentzian
functions were used. The simple Drude formula were used for $\sigma
_D(\omega )$. After subtracting the Drude and high frequency peaks in $%
\sigma (\omega )$, we obtained the $T$-dependent midgap component $\sigma
_{ms}(\omega )$[$=\sigma _I(\omega )+\sigma _{II}(\omega )$]. Note that the
polaron absorption and the inter-orbital transition between Mn$^{3+}$ sites
are assigned as Peak I and Peak II, respectively.

Peak I and II were fitted with two Gaussian functions, as shown in Fig. 5.
The solid circles represent the experimental $\sigma (\omega )$ after
subtracting $\sigma _{HS}(\omega )$, $\sigma _{CT}(\omega )$, and $\sigma
_D(\omega )$. The fitting results with the Gaussian functions could explain
the experimental data quite well. Using the integration of each Gaussian
function, we derived the $T$-dependent optical strengths, $S_I$ and $S_{II}$%
, for Peak I and Peak II, respectively. We also obtained the strength of the
Drude weight $S_D$ by integrating the corresponding Drude peak.

\begin{figure}[tbp]
\epsfig{file=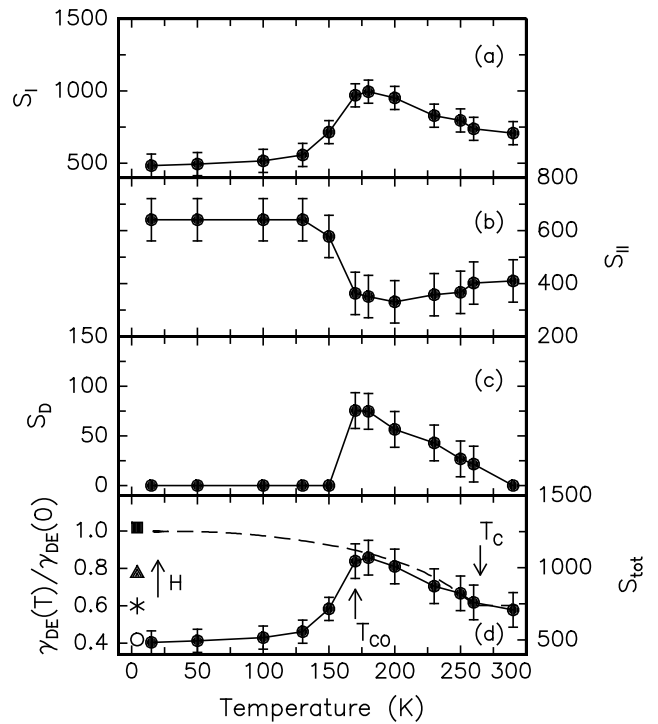,width=3.3in,clip=}
\vspace{2mm}
\caption{$T$-dependent optical strengths (a) $S_I$, (b) $S_{II}$, (c) $S_D$,
and (d) $S_{tot}$. [All units are $\Omega ^{-1}$cm$^{-1}$eV.] In (d), the
dashed line represents the prediction from the $T$-dependent double exchange
bandwidth. The open circle, the asterisk, the solid triangle, and the solid
square represent the $S_{tot}$ at 0, 12, 13, and 17 T, respectively.}
\label{Fig:6}
\end{figure}

The $T$-dependences of $S_I$, $S_{II}$, and $S_D$ are displayed in Figs.
6(a), (b), and (c), respectively. The total spectral weights due to polaron
absorption, $S_{tot}$($=S_I+S_D$), are also plotted in Fig. 6(d). [All units
are $\Omega ^{-1}$cm$^{-1}$eV.] With decreasing $T$, $S_I$ starts to
increase below $T_C$ and abruptly decreases below $T_{CO}$. The $T$%
-dependence of $S_{II}$ is nearly opposite to that of $S_I$. The $T$%
-dependence of $S_D$ is similar to $S_I$, but becomes zero for $T<T_{CO}$.
And, $S_{tot}$ starts to increase below $T_C$ and abruptly decreases below $%
T_{CO}$. When the sample becomes ferromagnetic below $T_C$, it becomes
metallic. Then, the polaron hopping contribution $S_I$ and the free carrier
contribution $S_D$ should increase due to the alignment of {\it t}$_{2g}$
spins. Due to the increase of metallicity, the lattice distortion of the MnO$%
_6$ octahedron will become reduced, resulting in decrease of $S_{II}$. When
the sample becomes antiferromagnetic below $T_{CO}$, the polaron hopping
requires more energy and $S_I$ should decrease. The reduction of the
metallicity makes $S_{II}$ increase and $S_D$ become zero very rapidly.
These temperature dependences are explained in the schematic diagram in Fig.
4.

According to the polaron picture, the $T$-dependence of $S_{tot}$ can be
explained more quantitatively. In (La,Pr)$_{0.7}$Ca$_{0.3}$MnO$_3$, whose
ground state is a 3-dimensional ferromagnetic metal, it was found that $%
S_{tot}$ could be scaled with the $T$-dependent double exchange bandwidth $%
\gamma _{DE}$ $(T)$:\cite{kim_lpcmo}

\begin{equation}
\gamma _{DE}=<\cos (\theta _{ij}/2)>,
\end{equation}
where $\theta _{ij}$ is the relative angle of neighboring spins and $<>$
represents thermal average in the double exchange model.\cite{kubo} This
scaling behavior was explained in a model by R\"{o}der {\it et al}.\cite
{roder} where the double exchange and the Jahn-Teller polaron Hamiltonian
were taken into account. The dashed line shows $\gamma _{DE}$ $(T)$ for the
3-dimensional ferromagnet. Above $T_{CO}$, the agreement between $S_{tot}(T)$
and $\gamma _{DE}$ $(T)$ is quite good.

However, $S_{tot}(T)$ deviates from $\gamma _{DE}$ $(T)$ below $T_{CO}$.
This deviation might be explained by a strong suppression of polaron
absorption due to the {\it CE}-type antiferromagnetic ordering at the low
temperature region. In the {\it CE}-type configuration, the {\it e}$_g$
conduction electrons are allowed to hop along the ferromagnetically aligned
zigzag chains forming an effective 1-dimensional ferromagnet. In the
3-dimensional ferromagnet above $T_{CO}$, the polaron hopping is allowed to
six neighboring Mn sites with parallel spins. But, in the 1-dimensional
ferromagnetic chain, the polaron hopping is allowed only along the zigzag
chain. Therefore, the transition from the 3-dimensional ferromagnet to the
1-dimensional zigzag chain will strongly suppress the polaron absorption
near $T_{CO}$.

\subsection{Magnetic field dependent spectral weight changes}

With the fitting process used in Section IV. B, we obtained $H$-dependent
changes of $S_I$, $S_{II}$, $S_D$, and $S_{tot}$, at 4.2 K. Figure 7 shows
the results of such fittings, and all of the optical strengths show strong
hysteresis behaviors. During the field-increasing run, $S_I$ increases near
13 T and becomes nearly saturated above 14 T. During the field-decreasing
run, $S_I$ remains nearly the same down to 8 T and then abruptly decreases.
Note that the value of $S_I$ at $H=$ 0 T after the completion of one cycle
is larger than the initial value of $S_I$. The $H$-dependence of $S_{II}$ is
nearly opposite to that of $S_I$, but the $H$-dependence of $S_{tot}$ is
similar to that of $S_I$. Contrary to rather smooth changes of $S_I$, $%
S_{II} $, and $S_{tot}$, the change of $S_D$ is rather abrupt: the value of $%
S_D$ becomes nearly zero below 13 T for field-increasing run and below 7 T
for field-decreasing run. Qualitatively, the $H$-dependences of $S_I$, $%
S_{II}$, and $S_D$, are quite similar to the $T$-dependences of the
corresponding strengths below $T_{CO}$. These $H$-dependences can be
explained using the schematic diagram in Fig. 4.

\begin{figure}[tbp]
\epsfig{file=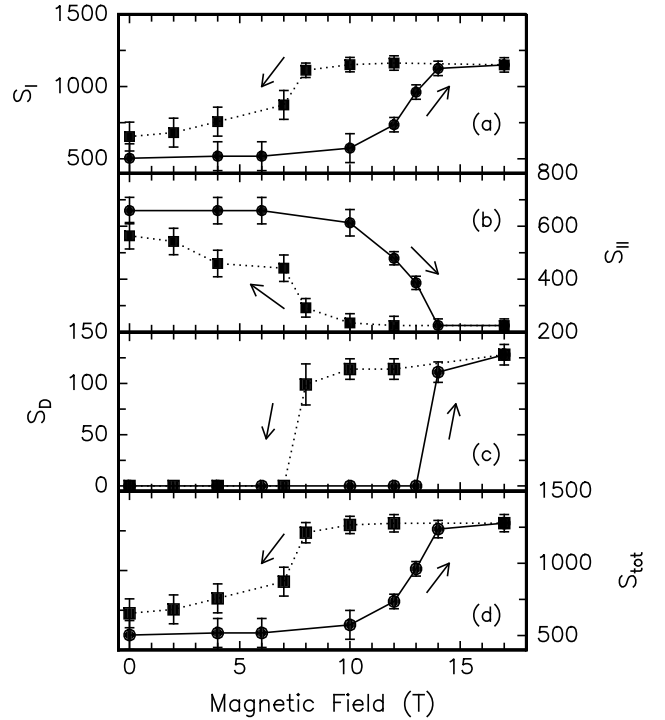,width=3.3in,clip=}
\vspace{2mm}
\caption{$H$-dependent optical strengths (a) $S_I$, (b) $S_{II}$, (c) $S_D$,
and (d) $S_{tot}$. [All units are $\Omega ^{-1}$cm$^{-1}$eV.] The solid and
the dotted lines are guides for eye for the field-increasing and the
field-decreasing runs, respectively. }
\label{Fig:7}
\end{figure}

Values of $S_{tot}$ at 4.2 K with various values of $H$ were shown in Fig.
6(d). The open circle, the asterisk, the solid triangle, and the solid
square represent values of $S_{tot}$ at 0, 12, 13, and 17 T, respectively.
With increasing $H$, $S_{tot}$ increases. At 17 T, it finally reaches the
value predicted by Eq. (2) for the 3-dimensional case. This result
indirectly supports the fact that the charge/orbital melted state might be a
3-dimensional ferromagnetic metal. Note that the value of $S_{tot}$ ($T=$
170 K, $H=$ 0 T) is by about 20 \% smaller than that of $S_{tot}$ ($T=$ 4.2
K, $H=$ 17 T). This experimental fact agrees with recent magnetostriction
measurement\cite{mahendiran} and transmission electron microscopy work\cite
{fukumoto} that there exist a local charge/orbital ordering even above $%
T_{CO}$.

\subsection{Behavior of dielectric constants}

To get a better understanding on the insulator-metal transition in Nd$_{1/2}$%
Sr$_{1/2}$MnO$_3$, we looked into the real part of a dielectric constant $%
\varepsilon _1$. The $T$- and the $H$-dependent $\varepsilon _1$ spectra are
shown in Fig. 8(a) and (b), respectively. In the insulating state at 15 K, $%
\varepsilon _1$ is positive and $d\varepsilon _1/d\omega \approx 0$. With
increasing $T$, $\varepsilon _1$ becomes slightly increased. Above 170 K, it
becomes abruptly decreased and $d\varepsilon _1/d\omega >0$, which are
consistent with typical responses of a metal. Note that the change in $%
\varepsilon _1$ is rather abrupt near the insulator-metal boundary. Such
interesting behaviors of $\varepsilon _1$ can be observed more clearly in
the $H$-dependence. Up to the insulator-metal phase boundary ($\sim $ 13 T), 
$\varepsilon _1$ is increased and then abruptly decreased. The changes of $%
\varepsilon _1$ and $\sigma $ at 100 cm$^{-1}$ under various $H$ are shown
in Fig. 9(a) and (b), respectively. These figures also show strong
hysteresis behaviors. The solid circles and the solid squares represent data
during the field-increasing and the field-decreasing runs, respectively. It
is clear that the abrupt change in $\varepsilon _1$ occurs near the
insulator-metal transition. Note that $\varepsilon _1$ becomes large as the
transition region is approached both from the insulating and from the
metallic sides.

\begin{figure}[tbp]
\epsfig{file=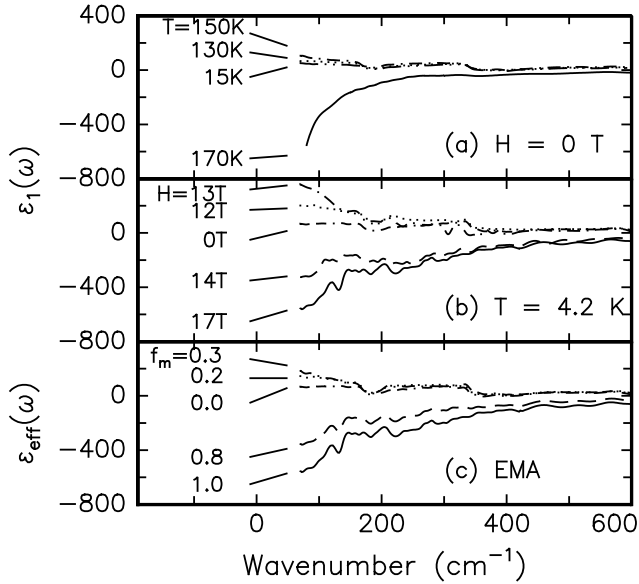,width=3.3in,clip=}
\vspace{2mm}
\caption{(a) $T$- and (b) $H$-dependent $\varepsilon _1$. (c) EMA results of 
$\varepsilon _{eff}$. In (c), $f_m$ represents the ferromagnetic metal
volume fraction.}
\label{Fig:8}
\end{figure}

The divergence of $\varepsilon _1$ near the insulator-metal transition has
appeared in numerous models. According to the Herzfeld criterion,\cite
{herzfeld,rama} valence electrons are considered to be localized around
nuclei and contribute to atomic polarizability. Near the insulator-metal
transition, the polarizability diverges, so $\varepsilon _1$ should also
diverge. Above the transition, the restoring force of the valence electron
vanishes, resulting in free carriers. Another is the Anderson localization
model.\cite{rama} The polarizability of a medium is proportional to square
of localization length. Since the localization length diverges near the
insulator-metal transition, $\varepsilon _1$ should diverge. [However, it is
clear that the insulator-metal transition in Nd$_{1/2}$Sr$_{1/2}$MnO$_3$ is
not induced by disorder.]

Note that both of the above microscopic models deal with $\varepsilon _1$
mainly in dc limit, so our infrared data cannot be well explained. And, it
was found that VO$_2$ films experienced an insulator-metal transition of the
first order nature around 70 $^{\circ }$C, and that their mid-infrared
properties could be explained by a composite medium model which takes into
account the evolution of domain growth during the first order phase
transition.\cite{choi} In the composite medium model, the increase of $%
\varepsilon _1$ near the insulator-metal transition can be interpreted as a
dielectric anomaly related to percolation. Therefore, we decided to apply
the effective medium approximation (EMA), which is a composite medium model
predicting a percolation transition.

\begin{figure}[tbp]
\epsfig{file=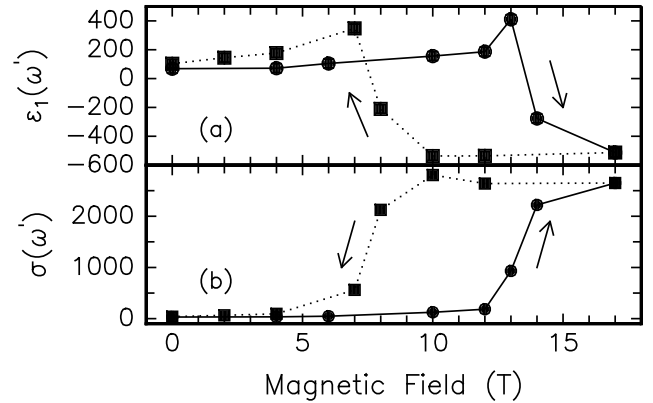,width=3.3in,clip=}
\vspace{2mm}
\caption{$H$-dependences of (a) $\varepsilon _1$ and (b) $\sigma $ at $%
\omega ^{^{\prime }}$ $=$ 100 cm$^{-1}$. The solid and the dotted lines are
guides for eye for the field-increasing and the field-decreasing runs,
respectively. }
\label{Fig:9}
\end{figure}

\subsection{Percolative phase transition}

In EMA, it is assumed that individual grains, either metallic or insulating,
are considered to be embedded in a uniform background, i.e., an ''effective
medium'' which has average properties of the mixture.\cite{choi,noh2} A
self-consistent condition such that the total depolarization field inside
the inhomogeneous medium is equal to zero leads to a quadratic equation for
an effective dielectric constant $\widetilde{\varepsilon }_{eff}$,

\begin{equation}
f_i\frac{\widetilde{\varepsilon }_i-\widetilde{\varepsilon }_{eff}}{%
\widetilde{\varepsilon }_i+2\widetilde{\varepsilon }_{eff}}+f_m\frac{%
\widetilde{\varepsilon }_m-\widetilde{\varepsilon }_{eff}}{\widetilde{%
\varepsilon }_m+2\widetilde{\varepsilon }_{eff}}=0\text{ ,}
\end{equation}
where $\widetilde{\varepsilon }_i$ and $\widetilde{\varepsilon }_m$
represent the complex dielectric constants of the insulating and the
metallic Nd$_{1/2}$Sr$_{1/2}$MnO$_3$ phases, respectively. And $f_i$ and $%
f_m $($=1-f_i$) represent volume fractions of the insulating and the
metallic domains, respectively. In EMA, the percolation transition occurs at 
$f_m=1/3$.

To apply EMA, we assumed that $\widetilde{\varepsilon }_i$ and $\widetilde{%
\varepsilon }_m$ could be represented by the experimental complex dielectric
constants at 0 T and 17 T, respectively. And, we evaluated $\widetilde{%
\varepsilon }_{eff}$ for various values of $f_m$. The predictions of EMA are
shown in Fig. 8(c). If $f_m<1/3$ (i.e. in the insulating side), $\varepsilon
_{eff}$ increases when the insulator-metal transition is approached. If $f_m$
becomes larger than 1/3, the low frequency value of $\varepsilon _{eff}$
suddenly becomes negative. By comparing with Fig. 8(a) and (b), it is clear
that the EMA results can explain the $T$-dependent and the $H$-dependent $%
\varepsilon _1$ quite well. It should be noted that the percolation model
can also explain the increase of $\varepsilon _1$ near the insulator-metal
transition. Near the percolation, effective capacitive coupling between the
metallic clusters increase due to an increase of effective area and the
decrease of spacing between the metallic clusters. This increase of coupling
results in the increase of $\varepsilon _1$ near the percolation transition.
Therefore, it can be argued that the insulator-metal transition in Nd$_{1/2}$%
Sr$_{1/2}$MnO$_3$ occurs through a percolative phase transition.

Recently, there have been lots of studies on the phase separations in doped
manganites.\cite{yunoki,moreo,allodi,fath,uehara,heffner,booth} Our picture
of the percolative phase transition agrees with such phase separation. The
EMA calculation in Fig. 8(c) shows that the metallic domain can exist in the
insulating states, i.e., below 150 K without $H$, or below 13 T at 4.2 K.
And it also shows that the insulating domain can exist in the metallic
states of Nd$_{1/2}$Sr$_{1/2}$MnO$_3$. The origin of the phase separations
in doped manganites remains controversial. Some workers argue that the phase
separation comes from the electronic origin,\cite{uehara} and some workers
argue that it comes from sample inhomogeneity.\cite{heffner,booth} Further
studies are required to solve this issue clearly.

\section{SUMMARY}

We reported the temperature and the magnetic field dependent optical
conductivity spectra of charge/orbital ordered manganites, Nd$_{1/2}$Sr$%
_{1/2}$MnO$_3$. With variation of the temperature and the magnetic field,
the large spectral weight changes were observed up to 4.0 eV. These spectral
weight changes are discussed with polaron picture and local charge/orbital
ordering. Moreover, using the analyses of dielectric constants, we showed
that the melting of charge/orbital ordered states occurs through the
percolation in the ferromagnetic metal domains and that optical conductivity
should be explained by the two-phase coexistence picture between
charge/orbital ordered insulator and the ferromagnetic metal domains.

\acknowledgements
We acknowledge Professor J.-G. Park and Dr. K. H. Kim for discussion. We
also thank Dr. H. C. Kim and Dr. H.-C. Ri for help in magnetoresistance
measurements. This work was supported by the Korea Science and Engineering
Foundation through the CSCMR at Seoul National University and by Ministry of
Education through the Basic Science Research Institute Program No.
BSRI-98-2416. The work by Y. M. was supported by a Grandt-In-Aid for Science
Research from the Ministry of Education, Science, Sports and Culture, and
from PRESTO, JST. Part of this work was preformed at the National High
Magnetic Field Laboratory, which is supported by NSF Cooperative Agreement
No. DMR-9016241 and by the State of Florida.

\end{document}